\documentclass[letter,12pt]{article}

\usepackage{titling}
\usepackage{fullpage}

\usepackage{authblk}

\usepackage{amsmath}  
\usepackage{amsfonts} 
\usepackage{graphicx} 
\usepackage{epstopdf}
\usepackage{gensymb}
\usepackage{color}
\usepackage{soul}

\usepackage{textcomp}

\usepackage{setspace}
\usepackage{caption}
\usepackage{sectsty}
\sectionfont{\fontsize{15}{12}\selectfont}
\renewcommand{\thesection}{\Roman{section}} 
\renewcommand{\thesubsection}{\thesection.\Roman{subsection}}

\usepackage[backend=biber,style=nature, doi=false,isbn=false,url=false,eprint=false,date=year]{biblatex}
\AtEveryBibitem{
\clearfield{note}
\clearfield{addendum}
\clearlist{language}
}
\AtEveryCitekey{\clearfield{note}\clearfield{addendum}\clearlist{language}}
\newcommand{\note}[1]{{\color{black}{#1}}}

\graphicspath{ {./figures/} }
\usepackage[utf8]{inputenc}
\usepackage{lineno}

\begin{document}

\begin{titlingpage}

\title{\textbf{{\large Non-reciprocity and multibody interactions in acoustically levitated particle systems: A three body problem}}}


\author[1,2,*,\dag]{Brady Wu}
\author[1,2,\dag]{Qinghao Mao}
\author[2]{Bryan VanSaders}
\author[1,2]{Heinrich M. Jaeger}

\affil[1]{Department of Physics, University of Chicago, 929 E 57th St, Chicago, IL 60637, USA}
\affil[2]{James Franck Institute, University of Chicago, 929 E 57th St, Chicago, IL 60637, USA}
\affil[*]{To whom correspondence should be addressed; E-mail:  bwu34@uchicago.edu}
\affil[$\dag$]{These authors contributed equally}

\maketitle

\begin{abstract}
\note{In active fluids and active solids the constituents individually generate movement by each extracting energy from their environment or from their own source. Non-reciprocal interactions among these active constituents then enable novel collective behavior that often can be strikingly counterintuitive. However, non-reciprocity in these cases typically requires that the interacting bodies have different physical properties or it needs to be programmed explicitly into all pairwise interactions. Here we show that collective activity in a driven system can emerge spontaneously through multibody nonreciprocal forces, even if all bodies are individually non-active and have identical properties. We demonstrate this with as few as three identical spheres, acoustically levitated in air, which exhibit collective activity as they interact through non-pairwise forces: similar to the classic gravitational three-body problem, the interaction between two spheres depends sensitively on the relative position of the third sphere.  Non-reciprocity arises naturally from both near-field sound scattering and microstreaming forces among the spheres. The underdamped dynamics in air furthermore make it possible to go beyond collective center-of-mass propulsion or rotation and observe internal, engine-like reconfigurations that follow limit cycles. These findings open up new possibilities for self-assembly, where now multibody interactions not only determine the resulting structure but also drive the spontaneously emerging dynamics. }
\end{abstract}

\end{titlingpage}

\section{INTRODUCTION}
A fundamental pillar of physics is Newton's third law, namely, action equals reaction.
However a wide spectrum of out of equilibrium biological, chemical and physical systems exhibit behaviors that seem to violate this principle on a coarse-grained level \cite{zampetaki_collective_2021, ivlev_statistical_2015,nagy_hierarchical_2010,meredith_predatorprey_2020}.
\note{The lack of reciprocity implies that a system is non-conservative and therefore requires that energy is supplied either by sources within the constituents or externally from the environment.
Irrespective of the energy sources, one particularly interesting consequence of nonreciprocal forces is their ability to generate emergent collective motion in multi-particle systems.
In interacting self-powered robotic units and externally driven rotating colloids, nonreciprocal forces have been found to give rise to self-propelled defects, non-reciprocal solitons, and non-reciprocal flocking \cite{bililign_motile_2022,veenstra_non-reciprocal_2024,fruchart_non-reciprocal_2021}.

In all of these systems, whether active or externally driven, non-reciprocity has typically been introduced by design through picking pairwise interactions that are asymmetric, i.e. particles with different dielectric constant, sizes, heights \cite{sukhov_actio_2015,king_scattered_2025,ivlev_statistical_2015}, preferred rotation handedness \cite{bililign_motile_2022, fruchart_non-reciprocal_2021}, or even programmed directly into (asymmetric) interactions between robotic constituents \cite{brandenbourger_limit_2022,veenstra_non-reciprocal_2024}.
Here we demonstrate a fundamentally different yet complementary approach, where the non-reciprocity emerges naturally from multibody effects. 
Through experiments and simulations we show that this can generate spontaneous collective dynamics in systems of individually passive particles even if their physical properties are identical and their pairwise interactions are reciprocal.

To show this we acoustically levitate small spheres in air, operating in a regime where the balance between sound-induced attractive and repulsive forces establishes stable configurations of slightly separated particles in the levitation plane.  
For just two spheres, the steady-state separation closely follows theoretical predictions \cite{fabre_acoustic_2017,wu_hydrodynamic_2023} and the interaction is reciprocal.  
However, with three (or more) spheres the situation is fundamentally changed.
The third particle renders the interactions multibody in nature, where now the force between any two particles depends on the position of the third particle.
In this 3-body system non-reciprocal interactions appear as soon as the particle cluster configuration deviates from its highly symmetric stable state, an equilateral triangle, and thus the symmetry is broken by \emph{dynamic} changes in configuration.
These lower-symmetry configurations are driven, and prevented from relaxing, by non-reciprocal multibody interactions between the identical constituent particles.
This is in contrast to pairwise non-reciprocal forces due to scattered waves, which require non-identical particles \cite{king_scattered_2025}.
For the minimal 3-sphere system we find two types of such emergent behavior. 
One is self-propulsion, where deviations from the equilaterial triangle induce a net force on the cluster as a whole. 
Here the configuration-induced multibody forces generate propulsion similar to that of colloids with designed shape asymmetry \cite{ahmed_density_2016,wright_migration_2008}, however here symmetry can break down spontaneously in a random direction. 
The other are limit cycle oscillations among the particles comprising the cluster, which resemble engine-like motion with a spontaneously chosen handedness. 
This type of underdamped, inertial dynamics uniquely requires the low viscosity environment afforded by levitation in air and is in an under-explored regime of non-reciprocal active matter.}

In experiments supported by extensive simulation we map out the conditions under which this leads to spontaneous self-propulsion of the 3-particle system or to limit cycle oscillations. 
Our levitated particles exist in a regime of fluid flow 
with Stokes number $\approx$ 5 and Reynolds numbers in the range 0.1 - 4. 
Here the Stokes number $\Omega = \frac{1}{2}(D/\delta_{\nu})^2$ is given by the ratio of particle diameter $D$ and viscous boundary layer thickness $\delta_{\nu}$ and, following Ref.\cite{dombrowski_kinematics_2020},  the Reynolds number is taken as $Re=2(s/D)\Omega$, where $s$ is the amplitude of the sound oscillation. 
Both the observed self-propulsion and the limit cycles constitute new behavior not reported  previously and also not captured by existing theoretical models, including those for micro-swimmers at low or intermediate Reynolds numbers\cite{najafi_simple_2004,derr_reciprocal_2022,lippera_no_2019}.

Fig.~\ref{fig:fig1}(a) is a sketch of the experimental setup, which injects high-intensity ultrasound into an acoustic cavity formed by the air gap between an ultrasonic transducer on top and a reflecting surface below it. 
With the gap height set to half a sound wavelength $\lambda$, a standing sound wave generates primary acoustic forces \cite{king_acoustic_1934,gorkov_forces_1962} to overcome gravity and lift small particles into a horizontal levitation plane just below the pressure node, halfway up the gap.
Additionally, as a result of the geometry of the acoustic transducer, a small radial gradient in the acoustic field inside the cavity weakly confines particles to the central region of the levitation plane.

Interactions among particles within the levitation plane arise from two types of sound-induced secondary forces \cite{silva_acoustic_2014,lim_acoustic_2024}: attractive forces $F_{sc}$ due to sound scattered off neighboring particles, and repulsive forces $F_{st}$ due to steady micro-streaming that originates from the interaction of a rapidly oscillating viscous fluid (air in this case) with the surface of levitated, effectively stationary objects.
The micro-streaming flow moves outward from the equators of particles in the levitation plane, and so establishes a short-ranged repulsion.
When this balances the attractive scattering force, a stable steady-state particle separation is enabled.
The experiments operate in the Rayleigh scattering regime, where the particle diameter $D$ is much smaller than the sound wavelength $\lambda$, and we furthermore focus on the extreme near-field limit, where the particle separation is also of the order of $D$.
In this limit all interactions between closely spaced particles arise solely from the interplay of secondary acoustic forces since the abovementioned radial confinement cannot provide sufficient force on the scale of a few particle diameters.

For two identical spheres the magnitudes of these forces depend only on the center-to-center distance $r$ and are known to scale as \cite{fabre_acoustic_2017,wu_hydrodynamic_2023}\\
\begin{equation}
    F_{sc}\propto-\frac{E_0D^6}{r^4},
\end{equation}
\begin{equation}
    F_{st}\propto E_0D^2~\exp(-(r-D)/\delta_{\nu}).
\end{equation}
Here $E_0\equiv p_{ac}^2/(2\rho_0c^2)$ is the sound energy density in the cavity, $p_{ac}$ is the sound pressure amplitude, $\rho_0$ is the density of the fluid (air), and  $c$ is the sound speed. 
Since the characteristic thickness $\delta_{\nu} = (2\nu/\omega)^{1/2}$ of the viscous boundary layer surrounding the particles depends only on the sound frequency $\omega$ and the kinematic fluid viscosity $\nu$, one can change the relative strength of $F_{sc}$ and $F_{st}$, and thus control the steady-state separation $r_{ss}$, simply by changing the particle size $D$ \cite{wu_hydrodynamic_2023}.
Note that the above expressions hold strictly only for two isolated, identical spheres. As will be shown below, this  fails to capture the rich physics if just one more sphere is added.

\section{EXPERIMENTAL SETUP AND METHODS}
\subsection{Experiment}

Acoustic waves were generated by a piezoelectric element and amplified by an aluminum horn, whose bottom surface formed the top boundary of an acoustic cavity trapping the particles. 
The bottom of the metallic horn had a concave radius of curvature $R=50$mm and was painted white to act as a back light for particle imaging when the cavity was illuminated from the side.
The concave geometry weakly confined the particles at the center of the acoustic trap.
A flat reflector plate formed the bottom surface of the acoustic cavity. 
To allow for imaging from below it consisted of a glass plate coated with transparent indium-tin-oxide. 
A heater tape, wrapped around the horn and regulated with a temperature controller, maintained the transducer at $35\pm 0.1^\circ$C for experimental repeatability.
The piezoelectric element was driven by applying a sinusoidal wave of peak-to-peak voltage $V_{pp}$ in the range 20-200V at a frequency close to the resonance frequency of the horn, $\omega_0/(2\pi)=34.870$kHz; this signal was generated by a function generator and amplified by a high-voltage amplifier (A-301 HV amplifier, AA Lab Systems).
The gap between the bottom of the horn and the reflector could be adjusted to $\lambda_0/2$ by moving the reflector plate with a translation stage, where $\lambda_0=9.8$mm is the wavelength associated with $\omega_0$. 
Throughout all experiments, we kept the gap height constant.

As particles we used polystyrene spheres (microparticles GmbH; material density $\rho=1000kgm^{-3}$, mean diameter $D=41.11\: \mu m$, standard deviation $\sigma(D)=0.5\: \mu m$).
The particles were added to the cavity with a surgical probe (tip diameter $\approx100\mu m$).
To reduce the effect of tribocharging, both the horn and reflector were grounded.
We neutralized any residual charges on levitated particles $\emph{in-situ}$ by briefly applying soft X-rays (Hamamatsu L12645 photo ionizer) at the start of an experiment. 
All experiments were performed in a humidity and temperature-controlled laboratory environment (45-50 \% relative humidity, 22-24$ ^\circ$C). 
We used de-ionized water to clean the bottom reflector and dried it with compressed air before and after each experiment. 
Videos were recorded with a high-speed camera (Vision Research Phantom T1340) at 6,000 frames per second.
The acoustic pressure in the cavity was measured with a laser-based acoustic sensor (Xarion Eta100 Ultra optical microphone).
In order to minimize the effect of air currents, the entire setup was enclosed within an acrylic box with dimensions much larger than the levitation region ($l\times w \times h=61\times 30 \times 46 \times  cm^3$).

\subsection{Finite Element Method Simulations}
We used COMSOL Multiphysics\textsuperscript{\tiny\textregistered} Software to perform finite element method (FEM) fluid simulations of particles levitating in the acoustic cavity. 
These 3D simulations accounted for all forces due to sound scattering as well as micro-streaming in the presence of the surrounding fluid (air). 
In order to obtain the time-averaged fluid properties, i.e.~velocity and pressure, we first solved the linearized Navier-Stokes equation using the \texttt{Thermoviscous Acoustics, Frequency Domain} module, which properly accounts for viscous effects and resolves the acoustic boundary layer. 
Performing the study in the frequency domain  ensured that the solution obtained was free from transients. 
Next, the second order time-averaged fluid properties were resolved by adding the necessary source terms (from the first order solution) to the \texttt{Laminar Flow} interface \cite{muller_numerical_2014}.
The total force on each particle was then obtained by summing  the viscous and pressure components, which we refer to as streaming and scattering components.
Particles and fluid were simulated in a rectangular box of height 10$D$ and width 40$D$ representing the acoustic cavity. 
The top and bottom walls were driven sinusoidally in phase with fixed displacement amplitude and frequency as in the experiments.
The side walls of the simulation box had slip boundary conditions to reduce finite-size effects. 
The minimal and maximal mesh sizes used around a particle were $0.6\mu m$ and $9.5\mu m$, respectively, to resolve viscous and streaming effects appropriately.
We controlled the COMSOL simulation with the \texttt{Matlab} interface to perform parameter sweeps on different configurations of particles.
The resulting forces for each configuration were then used as inputs to molecular dynamics simulations.

\subsection{Molecular Dynamics Simulations}
When the timescale of establishing a stable air flow field (including viscous flows) near the particles is much shorter than the timescale of the particle motion, we can conduct molecular dynamics (MD) simulations assuming steady-state conditions, i.e., the acoustic interaction forces felt by each particle are independent of their velocity through the fluid.
In this case, the forces on all three spheres in any given starting configuration can be calculated from an FEM simulation of non-moving particles, and input to a molecular dynamics (MD) simulation to calculate the next configuration.
Computing interaction forces from an FEM simulation at each time step of an MD simulation would be prohibitively time-consuming.
Rather, a grid of fixed particle configurations are pre-computed with FEM simulations and the forces on all particles are fit to an interpolation function \cite{berweger_molecular_1998, li_development_2022}.
Here, we first generated a batch of configurations, then fit the in-plane force components on each particle with a polynomial (for  details and a comparison between directly calculated and fitted forces see Appendix J and Fig.~\ref{fig:fit_comsol}).
With the fitting parameters in the polynomial generated beforehand, the calculation of the forces in the MD simulation was significantly accelerated by a factor of $10^6$.
We used a velocity Verlet algorithm to advance MD simulations in time, with a time step of $3.3\cdot10^{-4}$s and a total simulation time of $100-500$s to determine the final state for each initial condition.
The time step was reduced to $3.3\cdot10^{-5}$s to ensure stability when simulating with a larger viscosity $\nu>5.5\cdot10^{-5}m^2s^{-1}$. 


\section{RESULTS}
\subsection{Non-reciprocal and multibody acoustic interactions}
Figure \ref{fig:fig1}(b) shows the stable, equilateral triangle configuration for three levitated spheres of the same size (particle size dispersion can break the symmetry of the triangular configuration, see Appendix A).
To demonstrate that the interactions are not pairwise and become non-reciprocal, we performed simulations in which we vary the separation between particle 1 and 2, $r_{12}$, by moving the two particles in opposite directions along a circle of radius $r_{ss}$ with particle 3 kept fixed at the center  [Fig.\ref{fig:fig1}(c)].
As $r_{12}$ deviates from $r_{ss}$, a first observation is the strong effect this has on the force felt by particle 3 [Fig.\ref{fig:fig1}(d), movie 1]. 
While the force on particle 3 in a pairwise model can be expressed as $F_3(r_{13},r_{23})$ and thus should not depend on  $r_{12}$, here we clearly need to consider $F_3(r_{13},r_{23},r_{12})$.
Both scattering and microstreaming contribute in non-pairwise additive manner (see Appendix B for more details).

In the triangular configuration, when $r_{12}$ deviates from $r_{ss}$ the net force on all three particles no longer vanishes, a direct indication of non-reciprocity.
This can already be seen from the data in Fig.\ref{fig:fig1}(d) and is more directly shown in Fig.\ref{fig:fig1}(e), with contributions from scattering and micro-streaming apparent.
In particular, while reciprocal interactions obey $(\vec{F}_{1}+\vec{F}_{2})=-\vec{F}_{3}$, here as particle 1 and 2 come into near contact we find $\vec{F}_{1}\approx \vec{F}_{2}=\vec{F}_{3}$, implying spontaneous self-propulsion of the 3-particle `trimer’ configuration as a whole (in this example in the positive y-direction). 
The direction of this self-propulsion is reversed if $r_{12}$ is slightly larger than $r_{ss}$. However, because the streaming contribution is much weaker in this case, the magnitude of the net force on the trimer is significantly smaller.

Fundamentally, this multibody and non-reciprocal nature stems from the fluid that mediates the interactions between the particles; at each position, the fluid properties, such as its local pressure and velocity, depend on the boundary conditions, i.e., the position of every particle.
As a consequence, not only are the dynamics different for three particles, but also their stable separation $r_{ss}$ is slightly larger than for two (see Appendix C).
In fact, the interactions are multibody and non-reciprocal even when the three particles are arranged in a line (see Appendix D).
Multibody effects in other contexts often involve relative motion between the fluid and the objects in the directions of interest, giving rise to a viscous drag contribution\cite{yeo_collective_2015,janosi_chaotic_1997}.
By contrast, here the hydrodynamic coupling between particles is due to the oscillation of air normal to the levitation plane, which does not explicitly break symmetry in the levitation plane where particle motion takes place.

In the experiments we can self-assemble three particles into the stable trimer configuration and then excite an internal vibration to generate unbalanced forces that lead to global translation.
To this end we excite vibrations by parametric pumping, i.e.~amplitude modulating the air energy density $E_0$ at a frequency that matches the resonance frequency of an oscillation mode\cite{dolev_noncontact_2019}.
Figure~\ref{fig:fig1}(f) (bottom) depicts the resulting global translation; the gray lines trace the trajectory for the three particles, and the color denotes the temporal evolution.
Initially, the (symmetric) breathing mode is excited with small amplitude and therefore negligible unbalanced forces.
However as the oscillation grows the vibration becomes asymmetric, and the trimer with each cycle moves along the axis connecting the center of mass and particle 3 (as labeled in Fig.~\ref{fig:fig1}(c)), which we denote as the +y direction (Movie 2).
The propulsion results from the stronger net force in $+y$ direction when particles 1\&2 are in close proximity.
For $r_{12}>r_{ss}$ the total force on the system is in the $-y$ direction, but is significantly weaker.
Consequentially, for large enough oscillation amplitudes the time-averaged force will be in the $+y$ direction. 
This symmetry breaking is needed to generate a time-averaged propulsion force.
We emphasize that, for \emph{reciprocal} interactions, vibration modes are decoupled from translation and therefore the vibration-induced self-propulsion seen here is a direct demonstration of non-reciprocity.

Figure~\ref{fig:fig1}(e) reveals that the most pronounced self-propulsion occurs when particles 1\&2 are near contact. 
Propulsion would be maximized if these particles could be held at that separation (rather than oscillating).
In the experiments we can approximate the near-contact configuration by bringing two of the particles into actual contact, which reduces the thrust slightly\note{, but nevertheless allows us to directly explore an extreme regime of non-reciprocity from multibody interactions.}
To this end we induce collisions between two closely spaced particles via parametric pumping.
When these two particles stick due to (van der Waals) contact cohesion, a rigid dimer is formed that can be combined with a third, single particle [Fig.~\ref{fig:fig1}(g)].
Initially well separated, neither the dimer nor the single particle exhibit any self-propulsion.
However, once scattering interactions have attracted them into close proximity, they self-assemble into a stable configuration that translates spontaneously (see Movie 3).
In our experiments the distance traveled by the self-propelled trimer is limited by the very shallow radial confinement within the x-y levitation plane.
In Figs.~\ref{fig:fig1}(f) and (g) the self-propulsion speed therefore decreases as the trimer moves against this confinement, and motion eventually stops where the thrust balances the confining force.

While our experiments operate in a particle size regime where micro-streaming makes a dominant contribution to the non-reciprocity of the interactions, recent simulations have indicated that sound scattering alone can also drive spontaneous propulsion and rotation of suitably configured particle assemblies \cite{st_clair_dynamics_2023,king_scattered_2025}.
Conversely, earlier analytical work \cite{nadal_asymmetric_2014} predicted that ultrasound-induced micro-streaming alone should be able to generate self-propulsion of individual particles if they are sufficiently non-spherical.  
Such propulsion was observed when arrow-shaped nano-rods were acoustically levitated in water \cite{ahmed_density_2016,wright_migration_2008}. 
However, subsequent theoretical work \cite{lippera_no_2019} showed that, to leading order in asphericity and Reynolds number, net motion cannot arise from simple streaming flows, which led to the suggestion that additional rotational oscillation of a particle is required for the observed self-propulsion \cite{collis_autonomous_2017,nadal_acoustic_2020}.
As in our trimers the flow field inside the triangular particle configuration appears to play a major role, and as rotation of the 3-particle system is neither observed experimentally nor implemented in our simulations, our results demonstrate a new regime for sound-induced self-propulsion.
Notably, we find this propulsion even when the air oscillation amplitude $s$ is small ($s\sim 10^{-1}D$), and the system is in the Stokes flow regime ($Re=(s/D)\Omega \approx10^{-1}$), and therefore also in a regime different from that for swimmers at intermediate Reynolds numbers\cite{derr_reciprocal_2022,chen_self-propulsion_2024}.
Unlike models of swimmers that change their shape to propel (and therefore must break time-reversal symmetry with their motions)\cite{najafi_simple_2004}, here the geometry of the trimer can remain static during propulsion.
Instead, the thrust arises from the non-linear term $\vec{u}(\cdot \nabla)\vec{u}$ in the Navier-Stokes equation, where $\vec{u}$ denotes the fluid velocity.
This micro-streaming contribution to the thrust outweighs the contribution from sound scattering that points in the opposite direction (see Fig. 1(e) and Appendix B).

\subsection{Self-sustained limit cycles}

The low viscosity of air makes it possible to explore underdamped dynamics that cannot be observed with ultrasonic levitation in a liquid, where the particle dynamics typically are overdamped and inertial movements are quickly dissipated. 
For the dimer-plus-particle system we find that this enables a second type of spontaneously emerging dynamics driven by non-reciprocity: limit cycle oscillations around a stationary center of mass (Fig.~\ref{fig:fig2}(a) and Movie 4).  
We parameterize the system's internal oscillations by $r$, the distance between the dimer's center of mass and the particle, and $\theta$, the angle between the dimer's axis and $\hat{r}$ [Fig.~\ref{fig:fig2}(b)].
As the dimer and particle oscillate back and forth for millions of cycles without significant net propulsion, their movement follows a butterfly-like trajectory in $(r,\theta)$.
Experimental data for the example in Fig.~\ref{fig:fig2}(a) are shown in Fig.~\ref{fig:fig2}(c).
Also shown is the trajectory of a dimer-plus-particle simulated with our combined FEM/MD approach (red trace), which displays very similar limit cycle behavior.
We note that the simulated system very weakly self-propels during these oscillations, which we can also observe in some of the experiments (see Movie 4) but not reliably quantify, given the competing effects of weak confinement in the x-y levitation plane and residual experimental drift.

We observed two regimes of limit cycle behavior and will refer to them as regime I and II, and their onset is controlled by the magnitude of sound pressure ($p_{ac}$).
In regime I, as $p_{ac}$ increases, both the range and average of $r$ decrease, while the motion in $\theta$ stays nearly constant [Fig.~\ref{fig:fig2}(d)]. 
Regime II emerges as $p_{ac}$ is increased further, characterized by much smaller oscillation amplitudes than regime I and $r$ increasing with $p_{ac}$ [Fig.~\ref{fig:fig2}(e)].

Figure \ref{fig:fig2}(f) maps out, based on the experiments, how the overall dynamic behavior of the 3-particle system changes with increasing acoustic pressure.
Once $p_{ac}$ becomes large enough for levitation, initially only self-propulsion is observed.
In regime I, the dimer-plus-particle system can settle into either limit cycle (LC) or self-propulsion (SP) behavior, depending on the initial positions and relative velocities of the dimer and single particle as they approached each other during the self-assembly process.
This is also observed in the MD simulations (see below). 
Between regime I and II, we find a gap where the system will always self-propel, independent of the initial conditions.
Finally, in regime II, we find that the system will spontaneously start to oscillate, even when starting from an SP state.
In both regimes the limit cycle oscillation frequency $f$ increases with $p_{ac}$ in roughly linear fashion [Fig.~\ref{fig:fig2}(f)],
although this increase is much steeper in regime II.
Similarly, the average kinetic energy $E_{kin}$ increases with $f$ in both regimes, but the scaling is quadratic in regime I, as shown by the red dashed line [Fig.\ref{fig:fig2}(g)].

The above observations from the experiments indicate that the mechanisms driving the oscillations in the two regimes are different.
We note that our combined FEM and MD simulations, which assume Stokes flow for the contribution from the fluid and a steady-state acoustic field, can capture the behavior only for regime I, shown in Fig.~\ref{fig:sim_power}(a) and (b) in Appendix E.
With a particle-based Reynolds number $Re$ near 4 in regime II, the Stokes flow assumption is likely no longer valid, and furthermore the time scale of the oscillations approaches the response time of the acoustic field in the cavity (cavity quality factor $Q \sim$100).
Therefore, while a laminar flow assumption appears to be valid for regime I, that may not be the case for regime II, and  a time-dependent fluid simulation, appropriate for intermediate Reynolds numbers, would have to be implemented to reproduce the dynamics.
Another, more subtle finding is the decreasing distance $r$ between dimer and single particle with increasing $p_{ac}$ [Fig.~\ref{fig:fig2}(d)-(e)].
This is not recovered from our FEM fluid simulation.
It appears that with Reynolds number approaching $O(1)$ a fully developed streaming flow might be suppressed, thus weakening the strength of the repulsion.
Given these complications, in the following we restrict our analysis to regime I, the main features of which can be adequately captured by the simulation.

\subsection{The energy balance of limit cycles}
To address how the dimer-plus-particle system extracts energy to maintain the oscillations in regime I we now focus on building a minimal model.
We obtain the forces on each particle as a function of $r$ and $\theta$ from FEM fluid simulations.
It is convenient to express these as reduced radial force $\Tilde{F_r}=(\vec{F}_s/m_s - \vec{F}_d/m_d)\cdot \hat{r}$ and reduced torque $\Tilde{\tau}=\tau_s/I_s-\tau_d/I_d$, where $F_s$, $F_d$ ($\tau_s$,$\tau_d$) are the forces (torques) on the single particle (s) and dimer (d) respectively, while $m_s$, $m_d$ and $I_s$, $I_d$ are the mass and moment of inertia.
Here we take the difference between dimer and single particle acceleration to remove global linear or angular acceleration.
The resulting force field is shown in Fig.\ref{fig:fig3}(a). 
The forces on each particle are then utilized to perform an MD simulation, resulting in the red curve (same as in Fig.\ref{fig:fig2}(c), bottom).
At fixed $r/D$, the torque $\Tilde{\tau}$ as a function of $\theta$ is shown in Fig.\ref{fig:fig3}(b).
Around the stable position $\theta/\pi=1/2$, the response is linear and similar to a spring-like interaction.
However, the spring constant depends on $r/D$: at close approach, the restoring force in $\theta$ direction is nearly four times stronger than when dimer and single particle are far apart.
By scaling with $\exp(-r/\delta)$ and writing $\tilde{\tau}=-k_{\theta}(\theta-\pi/2)\exp(-(r-r_{ss})/\delta)$, where $k_{\theta}$ is the spring constant at $r_{ss}$, we can collapse these data (inset to Fig.\ref{fig:fig3}(b)).
This $\tilde{\tau}$ has a form similar to the acoustic streaming force $F_{st}$, the force primarily responsible for the coupling between $r$ and $\theta$. 

The coupling between $\theta$ and $r$ energizes the limit cycles. 
By adjusting the effective spring constant, the system loses less energy moving away from $\theta=\pi/2$ and gains more energy on the way back by decreasing $r$ and effectively stiffening the spring.
This is a form of parametric pumping, albeit self-organized by the system rather than externally imposed.

Figure \ref{fig:fig3}(c) shows the torque $\tau_{exp}$ estimated from experimentally obtained $\ddot{\theta}$ and $\dot{\theta}$ data (see Appendix F). 
To compare plots for different sound pressure levels, $\tau_{exp}$ is normalized by the acoustic energy $E_0V_p$, where $V_p$ is the volume of one particle.
The time-averaged experimental trajectory (light green) is in close agreement with the simulation (red curve).
Tracking the torque throughout an oscillation cycle we can find the work that has been extracted from the sound energy in the cavity via non-reciprocal interactions:
the average, normalized work  $W_\text{in}/(E_0V_p)$ per cycle that is gained from the $\theta$ component is the area enclosed by the green curve.
Comparing these areas for $p_{ac}=914$Pa (Fig.~\ref{fig:fig3}(c), left) and $1287$Pa (right) shows that with increasing $p_{ac}$ the system trajectories become less efficient at extracting energy. 
Yet, the ranges of $\tau_\text{exp}/(E_0V_p)$ and $\theta$ in the loop remain roughly the same for different acoustic pressures, so the energy pumped into a limit cycle, $W_\text{in}$, is proportional to $eE_0$, where $e$ is an efficiency factor associated with a specific, given limit cycle. This gives $W_\text{in}\propto ep_\text{ac}^2$.

In the steady state, the work $W_\text{in}$ energizing the limit cycle is exactly compensated by the work $W_\text{out}$ lost through viscous dissipation.
We can obtain a scaling for $W_\text{out}$ with $p_{ac}$ in regime I by assuming simple Stokes drag with coefficient $\gamma$, so that $W_\text{out}=\sum_{i=1}^3\int_0^{1/f} \gamma\vec{v_i}\cdot \vec{v_i} dt \approx \gamma \langle v^2\rangle/f=2\gamma E_{kin}/(mf)$, where the sum is over the three particles.
With $f$ scaling approximately as $p_{ac}$ and $E_{kin}$ as $f^2$ in regime I [Figs.~\ref{fig:fig2}(e,f)], we find $W_\text{out}\propto p_{ac}$.
Since $E_0 \propto p_{ac}^2$ this gives $W_\text{out}/(E_0V_p) \propto 1/p_{ac}$, which matches the experimental data well (dashed line in Fig.~\ref{fig:fig3}(d)).

Since $W_\text{in}$ scales quadratically with $p_\text{ac}$, but $W_\text{out}$ scales linearly, the efficiency factor $e$ in $W_\text{in}$ has to be adjusted to maintain $W_\text{in}=W_\text{out}$ when $p_\text{ac}$ is varied.
Given a family of LC trajectories, each associated with a different $e$, 
this means that the limit cycle trajectory has to change its shape as $p_\text{ac}$ changes.
The trajectory with the highest pumping efficiency $e$ determines the lower bound of $p_{ac}$ for possible LC behavior (left gray dashed line), below which $W_\text{in}<W_\text{out}$ and LC dynamics is not possible.
Likewise, the trajectory that has the lowest $e$ determines the upper bound of $p_{ac}$ for possible LC behavior, above which $W_\text{in}>W_\text{out}$ (right dashed line) and only self-propulsion can be observed experimentally.
Between these limits the system dynamically evolves the trajectories so that it can maintain energy balance, i.e. by decreasing $e$.
Similar to how random initial conditions are attracted into stable limit cycle orbits in a van der Pol oscillator \cite{b_theory_1920}, here each observed LC trajectory is the attractor for a particular value of $p_{ac}$ where $W_\text{in}=W_\text{out}$.
When $p_{ac}$ changes, the corresponding energy-balanced trajectory for the new $p_{ac}$ becomes the attractor, absorbing the previous trajectory.
Similar trends are found in simulations, where the energy scaling arguments outlined above hold exactly by construction (see Appendix E, Fig.~\ref{fig:sim_power}(c) and (d) for details).

The above observation has interesting implications to limit cycles in other mechanical systems with non-reciprocal interactions. 
If the strength of the interaction can be scaled and the dissipation is in the form of $-\gamma v$ (Stokes' drag), the same argument above holds, and one can expect the system to dynamically change the trajectory in a similar manner.
On the other hand, if the dissipation has the form $-\gamma v^2$, then both $W_\text{in}$ and $W_\text{out}$ will scale with $E_0$.
In this case, $W_\text{in}=W_\text{out}$ for all values of $E_0$, and the trajectory will remain unchanged for different $E_0$.

\subsection{Effect of damping on the limit cycle dynamics}

Based on the above discussion, viscous dissipation is one of the key ingredients in stabilizing the limit cycle dynamics.
In our experiments, which levitate particles in air, this dissipation cannot be easily changed.
Therefore, we turn to MD simulation and vary the effective damping coefficient $\gamma^*=\gamma/\gamma_{air}$ from $10^{-2}$ to $4$, where $\gamma_{air}$ is obtained from a best fit of MD simulation data to the experiment at $P_\text{ac}=1000$Pa.
Here, we keep the Stokes number $\Omega = \frac{1}{2}(D/\delta_{\nu})^2 $ unchanged across all simulations, which does not modify the form of acoustic interaction and thus allows us to directly compare experimental realizations with different damping.
In practice, this requires tuning the sound frequency $\omega$ in concert with $\gamma^*$ to preserve the viscous boundary layer thickness $\delta_{\nu}=\sqrt{2\nu/\omega}$.

Fig.~\ref{fig:fig4}(a) shows four simulated trajectories where the single particle approaches the dimer always from the same initial position, but with different $\gamma^*$.
The coordinate system is fixed on the dimer, i.e., we track the motion of the single particle relative to the dimer. 
At the beginning of each simulation, the single particle is released with zero velocity at the position marked by the plus sign. 
As the dynamics evolve, we find that, with a damping mimicking the experimental conditions, $\gamma^*=1$ (red), the single particle moves around the dimer and develops a limit cycle. 
When the damping is stronger, $\gamma^*=1.55$ (blue), kinetic energy is damped out within a few cycles, and the system attains the self-propulsion state. 
With slightly lower $\gamma^*=0.70$(orange), the single particle oscillates in a random manner, occasionally jumping from on side of the dimer to the other. 
The simulation at this $\gamma^*$ fails to enter a stable LC for as long as 500s, suggesting that there does not exist a feasible LC trajectory in the whole phase space to maintain the balance between energy extraction and dissipation. 
Finally, for very low $\gamma^*=0.01$ (green), the single particle orbits around the dimer and then flies away, as the viscous drag is not strong enough to dissipate the kinetic energy generated by the multibody non-reciprocal interactions.

In order to determine the likelihood of ending up in one of these four states (SP, LC, random, or separated) we scan the initial position of the single particle across the upper right quadrant in Fig.~\ref{fig:fig4}(a), for each starting point marking the final state by the corresponding color [Fig.~\ref{fig:fig4}(b)]. 
For $\gamma^*=1$ we also include the experimental observations, where light blue and light pink indicate SP and LC, respectively. 
While zero velocity initial conditions are harder to find in the experiments, the observed instances were fully consistent with the simulation predictions.

The data in Fig.~\ref{fig:fig4}(b) highlight an extreme sensitivity to the initial conditions, but they also indicate an overall evolution as the damping strength is varied.
Varying $\gamma^*$ and counting the total area occupied by each color then summarizes the probability of entering one of the four states [Fig.~\ref{fig:fig4}(c)].

With increasing $\gamma^*>1$, the limit cycle region is gradually replaced by self-propulsion state, and at $\gamma^*=1.55$, limit cycle ceases to exist. 
With a decreasing $\gamma^*<1$, LC becomes more probable than the SP before $\gamma^*=0.7$, where the limit cycle is replaced by the random state. 
Appendix H demonstrates that this transition is sharp at $\gamma^*$ between 0.75 and 0.76.
Below $\gamma^*=0.1$, both the SP near the stable configuration point and the outer random state are shrinking and gradually replaced by the separation state, i.e. a cluster of three particles is harder to find when the damping is too low. 
We also tested initializing the simulation with a large separation distance to mimic a far-field scattering experiment. Fig.~\ref{fig:fig4}(d) shows the probability for landing in each state after the single particle being released between $10-14D$ away from the dimer with zero velocity, and the dimer having any initial orientation. The trends in the results are similar to those with the near-field initial conditions, except that the transitions are somewhat sharper. This demonstrates that the transitions in the dynamics do not depend much on initial conditions, therefore showing a general consequence of tweaking the energy balance by tuning the damping coefficients.

Our simulations show that only a narrow range of damping coefficients, $\gamma^*=0.76-1.55$, can support the observed limit cycle behavior. This highlights the unique access to underdamped dynamics provided by levitation in air, and it explains why similar behavior has not been seen in other experiments where the particles are usually submerged in a liquid, and the viscous damping is much higher.
However, the simulations also show that some viscous damping is required for a stable limit cycle. For $\gamma^*<$ 0.76 the random state emerges because the system fails to find a stable trajectory where $W_\text{in}=W_\text{out}$.

\section{CONCLUSIONS}
\note{These results break new ground conceptually. They establish a model system that demonstrates how multibody forces can induce the non-reciprocal interactions needed to drive spontaneous dynamical behavior with otherwise non-active particles. This is in contrast to typical nonreciprocal interactions invoked so far, which have been pairwise, mostly involved sets of active objects, and tended to be prescribed explicitly rather than emerge naturally as here. These non-reciprocal forces cannot be captured by recent machine learning techniques assuming pairwise interactions \cite{yu_learning_2023}. An important consequence of the multibody nature of the nonreciprocity is that it does not require symmetry breaking `by design’. It thus can occur in systems of identical particles where the symmetry breaking is a result of configurational changes.} 

While forces due to acoustic streaming and scattering both are non-reciprocal and compete with each other, at small Stokes number $\Omega$ we find that acoustic streaming has the dominant effect and produces time-averaged flow capable of generating global translation of a 3-particle cluster when its configuration deviates from an equilateral triangle. The non-reciprocal effects are particularly enhanced and enable the cluster to power its dynamics by extracting most energy from the sound field when two of the three identical particles come close, thus breaking configurational symmetry. 
\note{As a proxy for such asymmetric arrangement the experiments use a dimer formed by two adhering particles, but conceptually this is not required. }

The resulting sound-driven self-propulsion differs from previously proposed mechanisms for the small $\Omega$ limit in that no rotation around the axis of propagation is involved. 
With acoustic levitation in air, furthermore, new regimes of underdamped, inertial dynamics with limit cycle behavior are accessed, which cannot be observed in a liquid environment. 
Specifically, for the dimer-plus-particle configuration we observe sustained limit cycle oscillations whose frequency increases with sound pressure level.
In the experiments we find two oscillation regimes, a low frequency, low-Re regime in which the limit cycles can coexist with self-propulsion, and a higher frequency, intermediate-Re regime in which only limit cycle oscillations are observed. Simulations appropriate for Stokes flow are found to capture the behavior in the first regime, detailing how the likelihood of observing either self-propulsion or limit cycle oscillation depends on how the single particle approaches the dimer; however, they fail to capture the behavior in the second regime, where Re approaches values around 4. In both regimes the limit cycles do not emerge from direct external driving as with the classic driven, damped harmonic oscillator or from single-agent active forces. They also differ from the collective limit cycles in lattices or chains of micro-robots with active forces \cite{zheng_self-oscillation_2023,baconnier_selective_2022}. Instead, they derive from the non-reciprocal interaction among multiple agents. This is similar to robotic units with coupling specifically designed to be non-reciprocal \cite{brandenbourger_limit_2022}, but here emerges naturally as a consequence of micro-streaming.

Taken together, these findings demonstrate how a set of small particles can extract energy from a sound field to perform both global translation and internal, engine-like motion, \note{both of which represent emergent dynamics that cannot be captured by the current state-of-the-art models for pairwise non-reciprocal interactions.}
With only three spheres this minimal system represents a fruitful test bed for investigating the underdamped dynamics that are likely to also drive the collective behavior of larger numbers of particles interacting through non-reciprocal multibody forces. It furthermore suggests new possibilities to utilize acoustic levitation as a platform for self-organization of small particles into micro-robots that can be powered and manipulated by sound. 
\note{More generally, we expect that the system studied here exemplifies emergent dynamic behaviors observable across a wide range of systems comprising nominally passive particles or objects with pairwise reciprocal interactions as they become `activated’ by multibody non-reciprocal interactions.}

\printbibliography

\newpage


 \begin{figure*}[h]
    \centering
    \includegraphics[width=172mm]{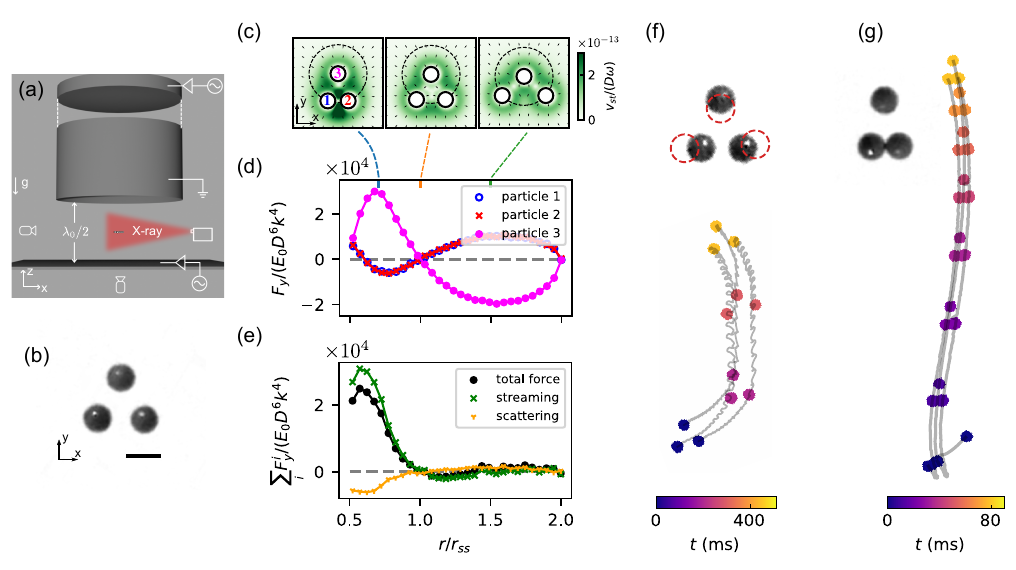}
    \caption{\textbf{Non-reciprocal and multibody acoustic interactions}. 
    (a) Sketch of the experimental setup.
    (b) Bottom view of  a levitating 3-particle cluster in its stable triangular configuration with center-to-center steady-state particle separation $r_{ss}$. Scale bar: 50$\mu m$.
    (c-e) Simulated deviation from the stable configuration, where particles 2 and 3 maintain a fixed distance to particle 3 (in center of dashed circle of radius $r_{ss}$) while their separation $r$ is varied: time-averaged streaming flow (c), net force in y-direction for each particle (d), and total force in y-direction (e). 
    (f) Oscillation induced by parametric pumping at 140Hz  generates time-averaged forces that propel the system forward.
    (g) Dimer and single particle self-assemble into a configuration that subsequently self-propels. In (f) and (g) snapshots of the particles taken from high-speed video are shown at the top, and a few select particle positions along the trajectories (gray lines) are color-coded according to elapsed time.  
    }
    \label{fig:fig1}
 \end{figure*}

 \begin{figure*}[h]
    \centering
    \includegraphics[width=172mm]{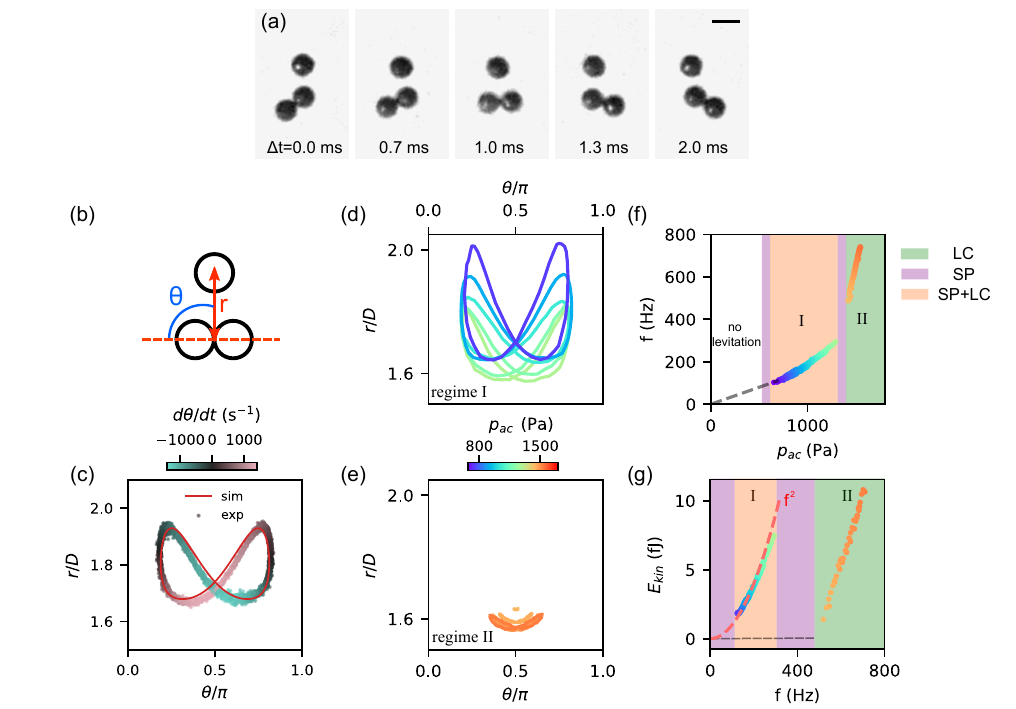}
    \caption{\textbf{Self-sustained limit cycles}. 
    (a) Snapshots of dimer and single particle exhibiting limit cycle dynamics. Scale bar: $50\mu m$.
    (b) Parameters $r$ and $\theta$ used to track limit cycle oscillation. 
    (c) Exemplary trajectory as function of $r$ and $\theta$, comparing experiment (data) and simulation (red line). 
    (d-e) Time-averaged trajectories from experiments at different acoustic pressure $p_{ac}$, indicated by color. 
    (f) Limit cycle oscillation frequency $f$ as function of $p_{ac}$.  
    Background color shading indicates the states observed (LC = limit cycle, SP = self-propulsion).
    (g) Time-averaged kinetic energy in the LC regimes as function of oscillation frequency. Red dashed line: $E_{kin}\propto f^2 $. The SP state has no internal motion (gray dashed line). The pressure color scale in (f)-(g) is the same as in (d)-(e).
    }
    \label{fig:fig2}
 \end{figure*}

 \begin{figure}[h]
    \centering
    \includegraphics[width=172mm]{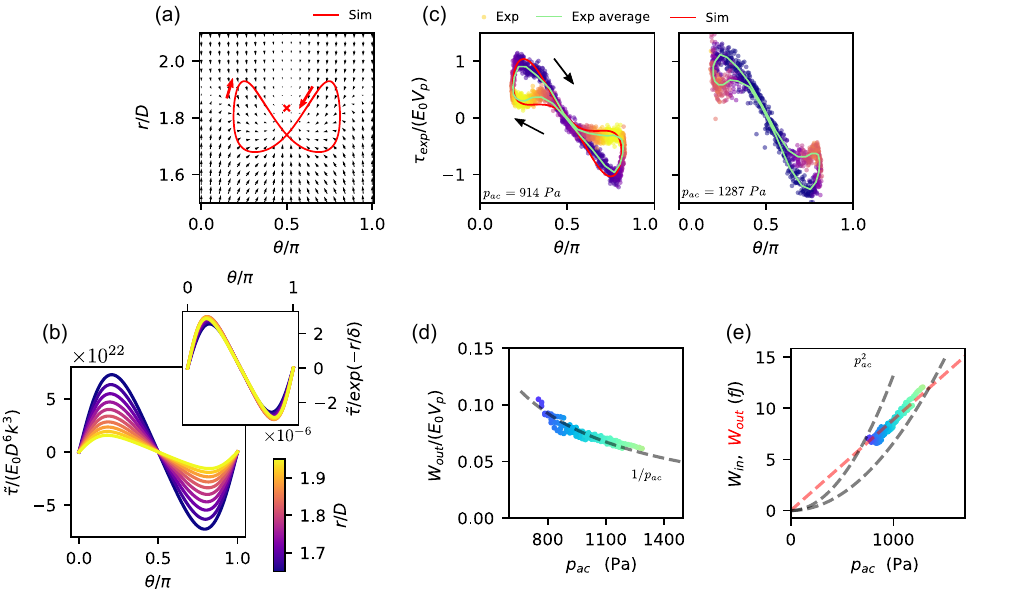}
    \caption{\textbf{Energy balance of limit cycles}. 
    (a) Reduced force field $(\tilde{\tau},\tilde{F}_r)$ in $\theta$ and $r$ from FEM simulation (arrows), used as input for MD simulation of trajectory (red line). 
    Arrow lengths are log-normalized.
    (b) Reduced torque $\tilde{\tau}$ as function of $\theta$ for different fixed $r$. The data can be collapsed by $\exp(-r/\delta)$, with $\delta=7.2\mu m$.
    (c) Torque ${\tau}_{exp}$ deduced from experiment, as function of $\theta$ for $p_{ac}=914$ (left) and $1287$Pa (right). Green line: time-averaged data. Red line: MD simulation. 
    (d) Normalized dissipated energy per cycle $W_\text{out}/(E_0V_p)$ as function of $p_{ac}$.
    (e) Limit cycles must balance energy pumped in $W_\text{in}$ and energy dissipated $W_\text{out}$ as $p_{ac}$ is changed.
    This is only possible where the family of LC trajectories of different pumping efficiency (grey, maximum and minimum) intersect the dissipation curve (red).
    }
    \label{fig:fig3}
 \end{figure}

 \begin{figure}[h]
    \centering
    \includegraphics[width=86mm]{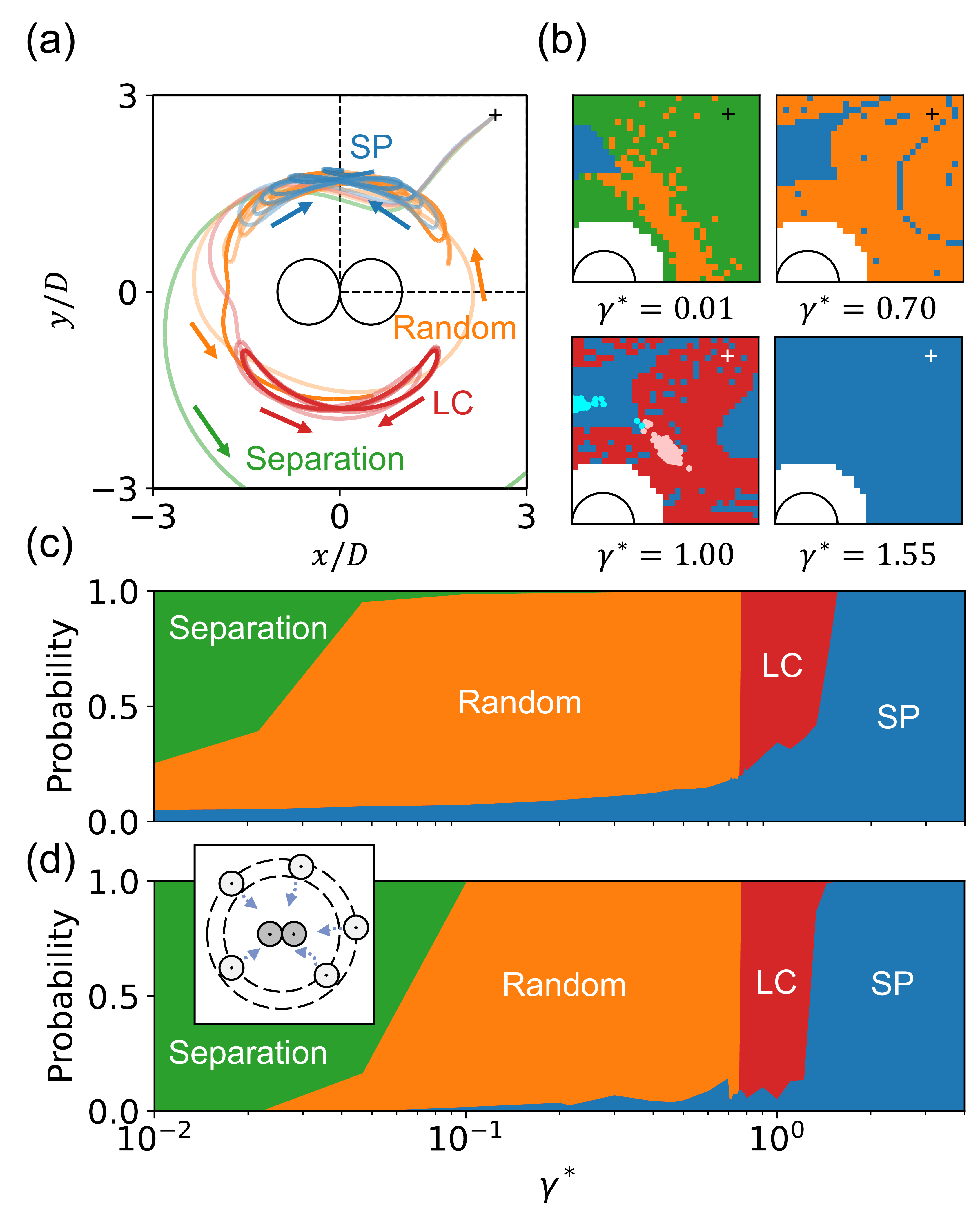}
    \caption{\textbf{Effect of damping on limit cycle dynamics}. 
    (a) Four representative trajectories of the single particle relative to the dimer, generated by MD simulations with the same zero-velocity initial condition starting from the location marked by the plus sign. Damping coefficient  $\gamma^*=0.01$ (green),  $0.70$ (orange), $1.00$ (red), and $1.55$ (blue).
    (b) Maps of final dynamic state after starting with zero-velocity from any location in the upper right quadrant of panel (a). Colors indicate SP (blue), LC (red), random (orange), separation (green).
    White regions are inaccessible due to  excluded volume.
    For $\gamma^*=1.00$ also data from experiment are shown, where light blue points end in a stable configuration, and light pink ones end in a limit cycle. 
    (c) Probability for each state shown in (b) as function of damping.
    (d) Same as (c) but starting the single particle from an initial position in the far-field. Inset: Sketch of different initial positions to obtain the data in (d).
    }
    \label{fig:fig4}
 \end{figure}

 \clearpage
 \newpage

\section*{Acknowledgements}
We thank Vincenzo Vitelli and Daniel S. Seara for insightful discussions. 
This research was supported by the National Science Foundation through award number DMR-2104733.
The work utilized the shared experimental facilities at the University of Chicago MRSEC, which is funded by the National Science Foundation under award number DMR-2011854. The research also benefited from computational resources and services supported by the Research Computing Center at the University of Chicago.

 \clearpage
 \newpage

 \setcounter{figure}{0}
\makeatletter 
\renewcommand{\thefigure}{A\@arabic\c@figure}
\makeatother


 \section*{APPENDIX A: TRIANGLE SHAPE OF THE TRIMER} 
 \label{apdx:triangle_shape}

\begin{figure}[h]
    \centering
    \includegraphics[width=86mm]{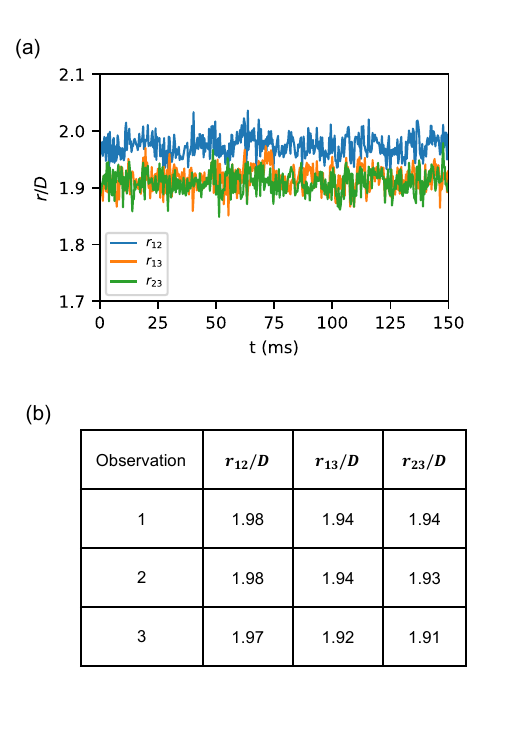}
    \caption{
    (a) Time traces of the separations between the three particles shown in Fig.~\ref{fig:fig1}(b).
    (b) Time-averaged separations obtained from three independent experiments.
    }
    \label{fig:triangle_length}
 \end{figure}
While the simulations show that the 3-sphere configuration for which all sound-induced forces balance is an equilateral triangle, the configurations observed in the experiments tend to deviate slightly from equilateral.  An example is given in Fig. \ref{fig:triangle_length}(a), where we plot the center-to-center particle separations $r_{12}$, $r_{13}$ and $r_{23}$ over time, and where the side of the triangle corresponding to  $r_{12}$ is measurably longer than the other two.
Similar length differences, all within about $\approx$3\%, were obtained from several independent experimental observations [Fig.~\ref{fig:triangle_length}(b)].
We attribute these  differences to slight variations in the sphere diameters, nominally around 0.5 $\micro$m. This corresponds to $\approx$1\% of $D$ and thus to a 2\% difference in Stokes number $\Omega \propto D^2$. Given that the steady-state particle separation depends sensitively \cite{wu_hydrodynamic_2023} on Stokes number  near $\Omega\approx 5$, which is the regime our experiments operate in, the slight particle size variations can easily produce the observed 3\% differences.

 \section*{APPENDIX B: SCATTERING AND STREAMING CONTRIBUTIONS TO FORCES IN THE TRIMER CONFIGURATION}

\begin{figure}[h]
    \centering
    \includegraphics[width=86mm]{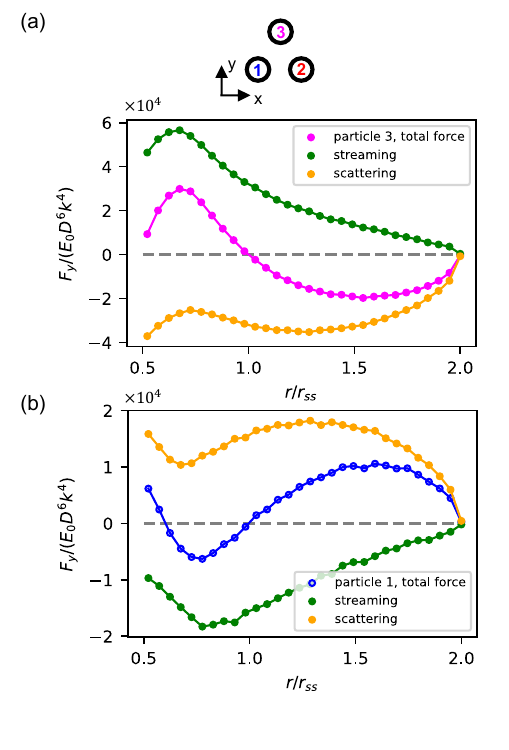}
    \caption{
    Forces in y-direction on particle 3 (a) and particle 1 (b), showing  the scattering and micro-streaming contributions. Data are from the simulation of the 3-particle configuration shown in Fig.~\ref{fig:fig1}(d).
    }
    \label{fig:particle13_breakdown}
 \end{figure}

Figure~\ref{fig:fig1}(e) in the main text shows the contributions from scattering and micro-streaming to the total force acting on the 3-particle trimer as a whole. 
Here we delineate the contributions from these two forces along the y-direction as they act separately on particles 1 and 3 (forces in the y-direction for particle 1 and 2 are the same due to symmetry; forces in x-direction cancel due to symmetry).
This is shown in Figs. \ref{fig:particle13_breakdown}(a) and (b).
For small $r$ both particles experience positive net forces in y-direction, but from different contributions.
For particle 3, the force in +y direction mainly comes from streaming, while for particles 1 and 2 the force in +y direction mainly originates from scattering.
Non-reciprocity is the strongest in the streaming contribution: when particle 1 and 2 are near contact, the magnitude of the streaming force on particle 3 is nearly five times larger than the one on particle 1 and 2, a result of the strong streaming flow in y-direction between the dimer particles, as shown in Fig.~\ref{fig:fig1}(c) of the main text.

 \section*{APPENDIX C: ADDITION OF  THIRD PARTICLE INCREASES STEADY-STATE SEPARATIONS}

\begin{figure}[h]
    \centering
    \includegraphics[width=86mm]{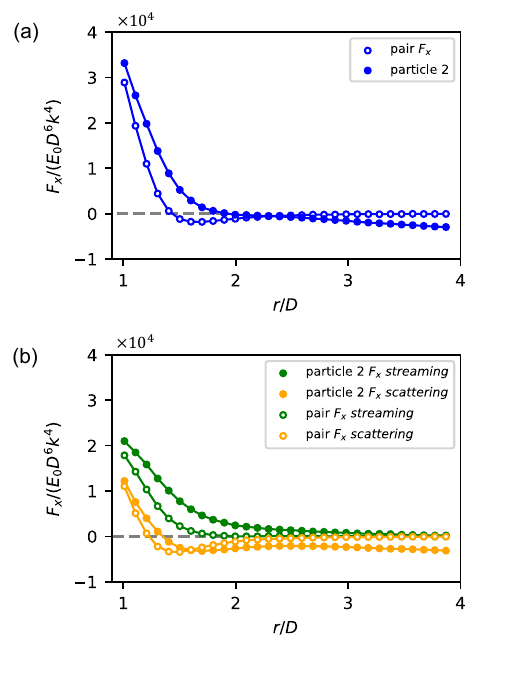}
    \caption{(a) Net force on particle 2 of a trimer in the +x direction, in comparison with the forces between particles in an isolated particle pair situated on the x-axis.
    (b) Micro-streaming and scattering contributions to the forces shown in (a). Data for both plots are from the simulation shown in Fig.~\ref{fig:fig1}(c).
    }
    \label{fig:fx_multibody}
 \end{figure}

Figures~\ref{fig:fig1}(c)-(e) demonstrate the multibody and non-reciprocal character of acoustic interactions, but only show the force in the y-direction.
Here we plot the force in the x-direction and compare it with a system of only two particles, highlighting the resulting change in steady-state particle separation.
Figure~\ref{fig:fx_multibody}(a) shows the force in +x direction on particle 2 of a trimer and on one of the particles in an isolated pair of particles situated along the x-axis (for the 2-particle system the force is reciprocal).
Most notably, the presence of the third particle increases the steady-state separation. 
Additionally, for increasing $r/D>2$, the attraction between the pair of particles decays, but $F_x$ increases on particle 2 in the trimer.
Note that in these simulations  $r_{23}$ is always kept at $r_{ss}$.
As $r/D$ increases to 3, the force on particle 2 is mostly due to particle 3, which is still attractive.
When $r/D\approx 4$, the three particles are arranged in a line.
See Appendix D for more information about interparticle forces in this configuration.

Figure \ref{fig:fx_multibody}(b) shows the relative contributions of the micro-streaming and scattering forces in the x direction.
The streaming component differs the most between the two- and three-particle scenarios.
Furthermore, the streaming force is significantly stronger when there are three particles, which is the driver for the increase in the trimer's particle separation.

 \section*{APPENDIX D: MULTIBODY AND NON-RECIPROCAL EFFECTS FOR LINEAR PARTICLE CONFIGURATION}
\begin{figure}[h]
    \centering
    \includegraphics[width=86mm]{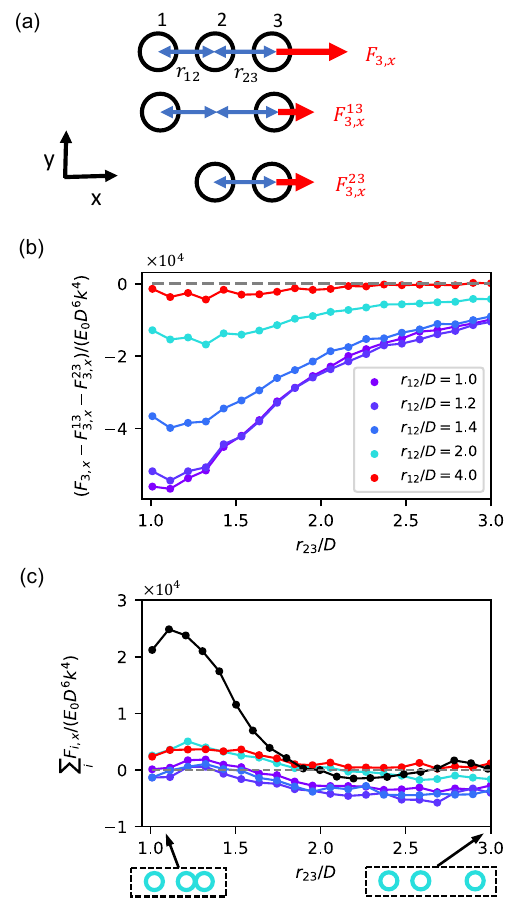}
    \caption{
    (a) Three levitated particles arranged in a line along the x-direction.
    $F_{3,x}$ is the net force on particle 3 in the x-direction resulting from the presence of the other two particles (top).
    $F^{i3}_{3,x}$ with $i=1,2$ is the net force in the x-direction on particle 3 when only particles 1 and 3 or 2 and 3 are present (middle, bottom).
    (b) $F_{3,x}-F^{13}_{3,x}-F^{23}_{3,x}$ as  function of $r_{23}$ for five different values of $r_{12}$ is shown.
    Deviations from the horizontal dashed line indicate non-pairwise interaction. 
    (c) Forces in x-direction summed over all three particles $\sum_i F_{i,x}$ as  function of $r_{23}/D$ and $r_{12}/D$. Deviations from the horizontal dashed line indicate a  non-zero total force and thus non-reciprocal interactions. Depending on whether particle 2 is closer to particle 1 or 3, this can lead to self-propulsion either in positive or negative x-direction, as shown by the bottom schematic.
    The thrust here is much weaker than the 3-particle system shown in Fig.~\ref{fig:fig1}(e), also shown here in black.
    }
    \label{fig:multibody_line}
 \end{figure}

We can further explore the nature of the acoustic interactions between three levitated particles by considering a configuration where they are arranged along a line.
The schematic of the simulation setup is shown at the top of Fig.~\ref{fig:multibody_line}. 
We vary the distance between particles 2 and 3, $r_{23}$, for five representative values of $r_{12}$ and extract the force on particle 3 in the x-direction, $F_{3,x}$.

To show that the interaction is multibody, let us first assume that the interaction is pairwise. 
Under this assumption, the total force on particle 3  is the sum of the forces due to particle 1 and particle 2, i.e., $F_{3,x}=F^{13}_{3,x}+F^{23}_{3,x}$.
To test this, we compare $F_{3,x}$, calculated by considering all three particles as described above, with  similar simulations to obtain $F^{13}_{3,x}$ and $F^{23}_{3,x}$ when only particles (1 and 3, or 2 and 3) are present.
If the interactions were pair-wise, $F_{3,x}-F^{13}_{3,x}-F^{23}_{3,x}$ would be zero (gray dashed line) for all values of $r_{23}$ and $r_{12}$.
However, Fig. 4A(a) shows that this is only true when $r_{12}/D=4$ and therefore only when particle 1 is weakly interacting with the rest.
This demonstrates that the interaction is multibody even when all three particles are positioned on the same line.

To demonstrate the non-reciprocity, we sum up the forces on all three particles to obtain $\sum_i F_{i,x}$ as a function of $r_{23}/D$ and $r_{12}/D$ [Fig.~\ref{fig:multibody_line}(c)].
We find that total force is nonzero almost everywhere and, in general, the total force is in the direction of the two particles that are closer to each other.
We  therefore conclude that the acoustic interaction is multibody and non-reciprocal for nearly all configurations of the 3-particle system as long as  the separations between the particles are in the range $r/D<4$.

\section*{APPENDIX E: SIMULATING LIMIT CYCLE IN REGIME I}
\begin{figure}[h]
    \centering
    \includegraphics[width=3.2in]{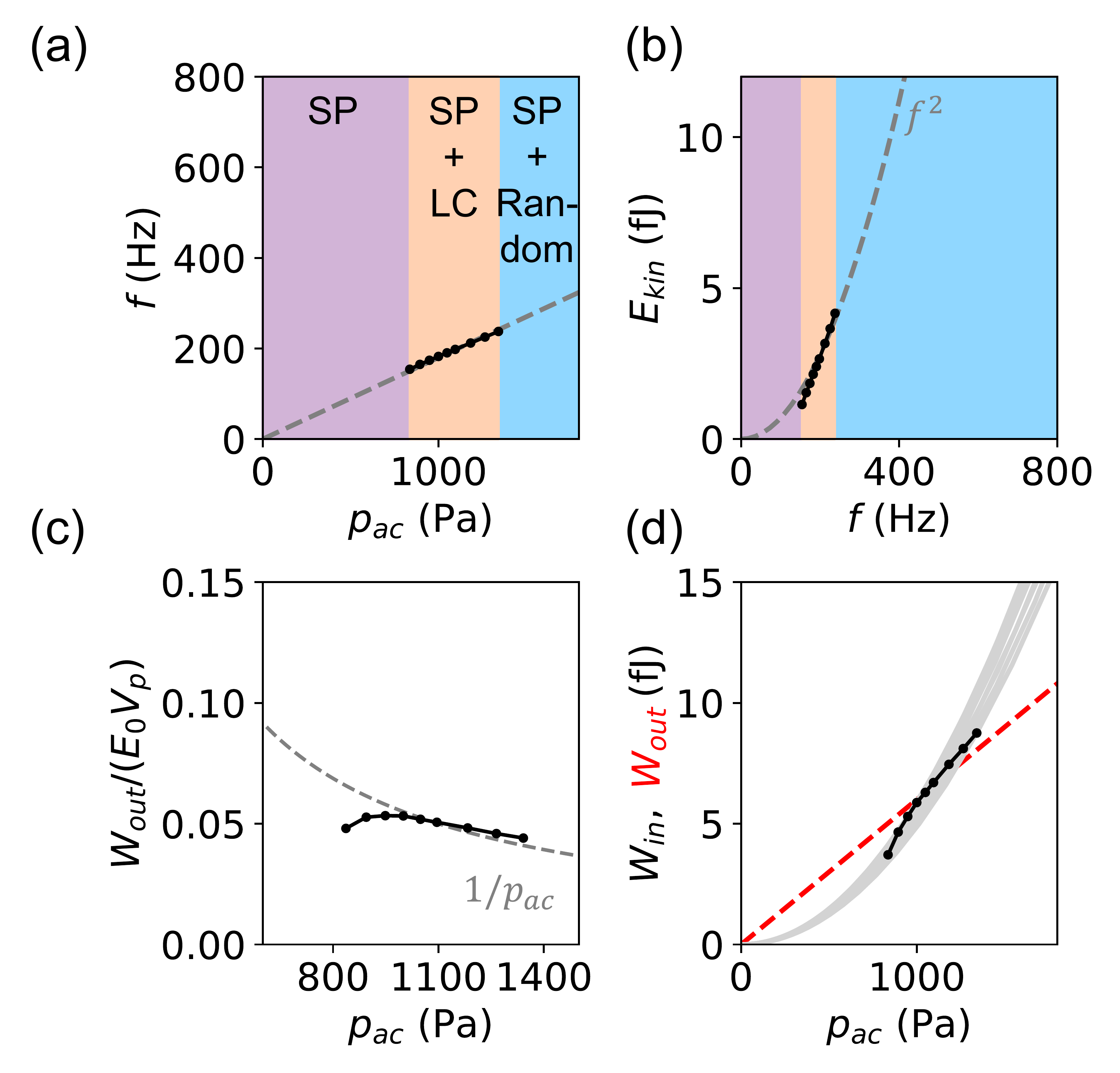}
    \caption{MD simulation of limit cycles in Regime I. (a) Time-averaged LC frequency increases linearly with $p_\text{ac}$. (b) Time-averaged kinetic energy grows quadratically with frequency. (c) Normalized $W_\text{out}$ decreases as a function of $1/p_\text{ac}$. (d) Black points show $W_\text{out}$ increases linearly with $p_\text{ac}$. Each light gray curve shows $W_\text{in}$ of one particular trajectory as a function of $p_\text{ac}$, and its crossover point with the dashed line indicates where $W_\text{in}=W_\text{out}$. The deviation of  $W_\text{out}$ from the straight line comes from the scaling deviation of $E_\text{kin}$ from $f^2$, where shape change in the trajectory is not captured by $f^2$.
    }
    \label{fig:sim_power}
 \end{figure}

After the parameters for interaction strength and damping used in the MD simulation have been calibrated by matching the simulated trajectory with the experimental one shown in Fig.~\ref{fig:fig2}(c), numerical investigation of limit cycle (LC) behavior in Regime I can be carried out. 
Figure~\ref{fig:sim_power} shows that the simulation  produces results that  qualitatively match those seen in the experiments (see Figs. 2(e)-(f) and Figs. 3(d)-(e)). 
Figures~\ref{fig:sim_power}(a) and (b) show the linear change of the LC oscillation frequency with sound pressure and the approximately quadratic increase of kinetic energy with oscillation frequency, while Fig.~\ref{fig:sim_power}(c) supports the experimental observation that the normalized work due to dissipation scales as $1/p_{ac}$, except for the lowest acoustic pressure levels.
Taken together, these simulation results lend further support to the notion that the beginning and ending of LC regime I is controlled by the energy balance $W_\text{in}=W_\text{out}$, as shown in Fig.~\ref{fig:sim_power}(d).

Similar to the simulations in Fig.~\ref{fig:fig4}, this simulation assumes an unchanged boundary layer thickness $\delta_\nu$, and thus does not capture the change in the average separation distance between dimer and single particle with increasing $p_\text{ac}$.
We speculate that this perhaps explains why the simulation predicts SP and random dynamics rather than only SP (as seen in the experiments) when $p_\text{ac}$ is increased beyond regime I. 

 \section*{APPENDIX F: ESTIMATION OF $\tau_{exp}$}
The experimentally observed $\ddot{\theta}$ includes both damping and interactions between dimer and the single particle.
We can write the equation of motion for $\theta$ as follows:
\begin{align*}
    I\ddot{\theta} &=\tau_{exp}-\gamma_{\theta}\dot{\theta},
\end{align*}
where $I$ is the moment of inertia and $\gamma_{\theta}$ is the damping in the $\theta$ degree of freedom.
$\dot{\theta}$ and $\ddot{\theta}$ can be experimentally measured. Therefore, we only need to estimate $I$ and $\gamma_{\theta}$ to obtain $\tau_{exp}$.
If we approximate the moment of inertia as $I=mD^2$, the experimental data 
gives $I_d=2.3mD^2$ and $I_s=3.5mD^2$ for the moment of inertia of the dimer and the single particle, respectively.
$\gamma_{\theta}$ can  be estimated from the Stokes drag on a single particle $\gamma=6\pi\mu (D/2+\delta_{\nu})$, where $\mu$ is the dynamic viscosity of air and $\delta_{\nu}\approx 12\mu m$ is the size of the viscous boundary layer.
The damping force on a single particle is $\gamma v$. 
Converting it to torque, we have $\gamma v D=\gamma \dot{\theta}D^2$.
Therefore, $\gamma_{\theta}=D^2 \gamma$.
We can then get $\tau_{exp}$ from 
\begin{align*}
    \tau_{exp}&=I(\ddot{\theta}+\gamma_{\theta}/I\dot{\theta}),
\end{align*}
where $\gamma_{\theta}/I=\gamma/m\approx 300$. 
We note that the value of $\tau_{exp}$ obtained with the above approximations is only an order of magnitude estimate. 
Nevertheless, we find good agreement by comparing the area enclosed by the green line in Fig.\ref{fig:fig3}(c) of the main text with the values in Fig.\ref{fig:fig3}(d).


\section*{APPENDIX G: FORCE FITTING}

\begin{figure}[h]
    \centering
    \includegraphics[width=5in]{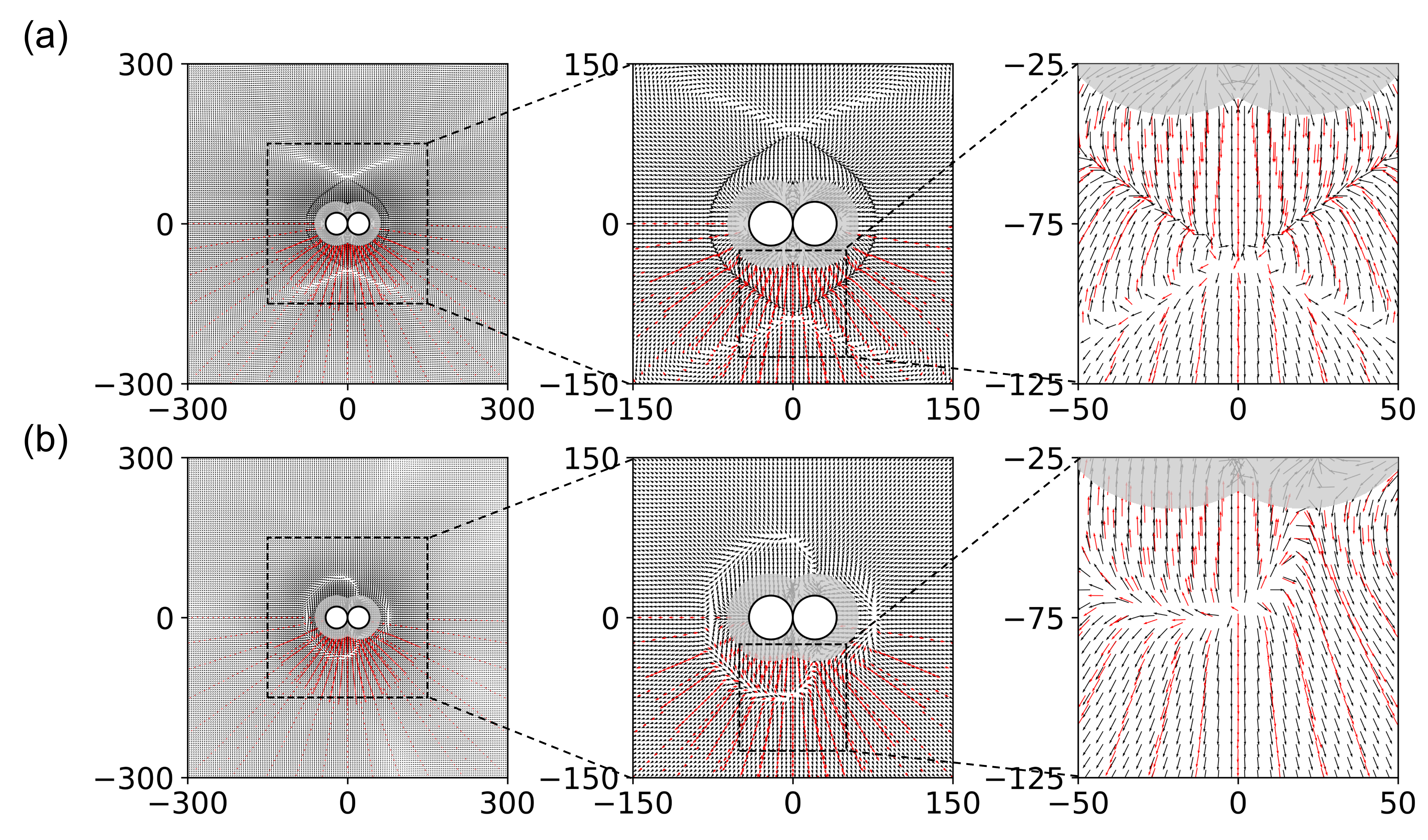}
    \caption{Fitting of forces between particles based on COMSOL simulation. Force map of (a) the single particle, and (b) the left particle in the dimer. All numbers indicate distances in units of microns. The second and third column shows a zoomed-in view of the area marked by the dashed black square in the left column. Red arrows: Results from direct FEM simulation. Black arrows: Fitting results. The arrow length are plotted in log scale. Gray area: impossible region where the single particle and the dimer are in contact. 
    }
    \label{fig:fit_comsol}
 \end{figure}

The fitting of forces from FEM simulations uses a polynomial consisting of terms ranging from $r^{-4}$ (for the long-range attractive part) to $r^{-10}$ (for the short-range repulsive part) and angular terms of $\sin(k\theta)$ and $\cos(k\theta)$ with integer $k$ from 1 to 9. 
Figure~\ref{fig:fit_comsol} compares the interpolated forces with forces calculated directly by FEM numerics. In both panels, we fix the position and orientation of the dimer, marked by the two particles in the center, and sweep the position for the single particle in both x and y directions. In Fig.~\ref{fig:fit_comsol}(a), each arrow shows the force on the single particle if the center of the single particle is at the arrow's position.
In Fig.~\ref{fig:fit_comsol}(b), the force on the left particle of the dimer is shown instead. The force on the right particle of the dimer can be calculated based on symmetry. The zoomed-in panels show that the interpolated forces are in good agreement with the ones obtained by direct FEM simulation.

\section*{APPENDIX H: TRANSITION BETWEEN RANDOM AND LC STATE CONTROLLED BY $\gamma^*$}

\begin{figure}[h]
    \centering
    \includegraphics[width=4.5in]{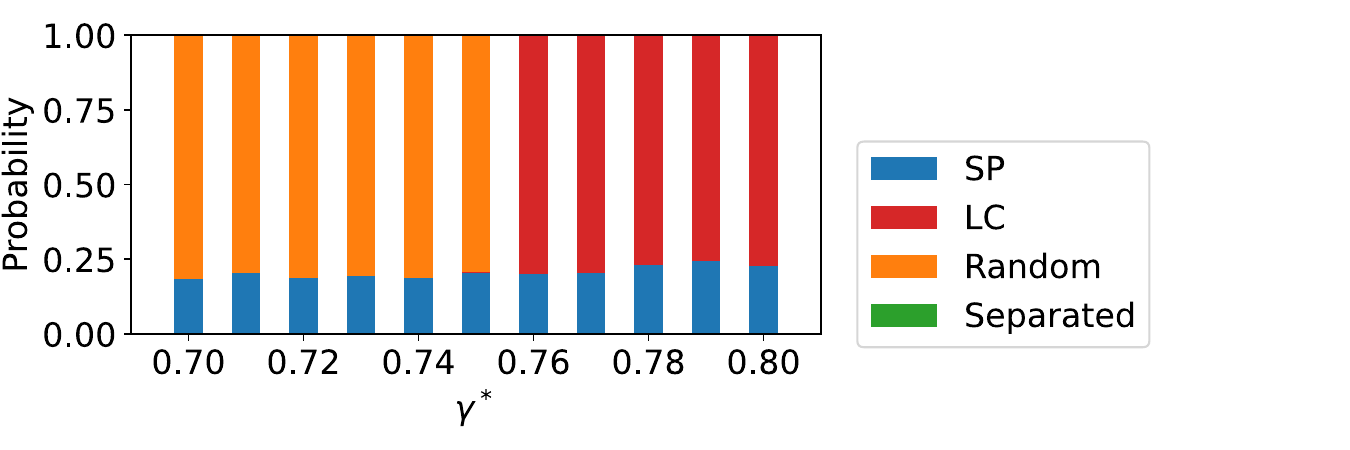}
    \caption{Probability for each state shown in Fig.~\ref{fig:fig4}(c) between $\gamma^*=0.7$ and $0.8$.
    }
    \label{fig:state_diagram_zoomin}
 \end{figure}

Figure~\ref{fig:state_diagram_zoomin} shows in more detail the probability for each state to occur when the dissipation is varied between $\gamma^*=0.7$ and $0.8$. We find that the transition between the random state and LC happens sharply between $\gamma^*=0.75$ and $0.76$. This suggests that $\gamma^*\approx 0.75$ is a threshold below which the system can no longer find a stable trajectory that satisfies $W_\text{in}=W_\text{out}$.

\end{document}